\documentclass[showpacs,preprintnumbers,amsmath,amssymb,aps,pra,groupedaddress]{revtex4-2}
\usepackage{multirow,diagbox}  
\usepackage{siunitx}
\usepackage[colorlinks,citecolor=blue,linkcolor=red,hyperindex,CJKbookmarks]{hyperref}
\usepackage{amsmath}
\usepackage{graphicx}
\usepackage{dcolumn}
\usepackage{subfigure}
\usepackage{mathrsfs}
\usepackage{amssymb}
\usepackage{amsfonts}
\usepackage{makecell}
\usepackage{color}
\newif\ifistoreview
\istoreviewtrue
\istoreviewfalse
\newcommand{\removeColor}{\textcolor{red}}
\newcommand{\addColor}{\textcolor{blue}}
\newcommand{\deleted}[1]{\ifistoreview\removeColor{\st{#1}}\else \fi}
\newcommand{\added}[1]{\ifistoreview\addColor{#1}\else #1\fi}
\newcommand{\replaced}[2]{\ifistoreview\added{#1}~\deleted{#2}\else #1\fi}
\begin{document}
\draft
\title{Quantum metric and metrology with parametrically-driven Tavis-Cummings \replaced{models}{model}}
\author{Jia-Hao L\"{u}, Pei-Rong Han, Wen Ning, Xin Zhu}
\author{Fan Wu}\thanks{E-mail: t21060@fzu.edu.cn}
\author{Li-Tuo Shen}
\author{Zhen-Biao Yang}\thanks{E-mail: zbyang@fzu.edu.cn}
\author{Shi-Biao Zheng}\thanks{E-mail: t96034@fzu.edu.cn}
\address{Fujian Key Laboratory of Quantum Information and Quantum\\
	Optics, College\\
	of Physics and Information Engineering, Fuzhou University, Fuzhou, Fujian
	350108, China}
\date{\today }

\begin{abstract} 
	We study the quantum metric in a driven Tavis-Cummings model, comprised of multiple qubits interacting with a quantized photonic field. The parametrical driving of the photonic field breaks the system's U(1) symmetry down to a ${\rm Z}_2$ symmetry, whose spontaneous breaking initiates a superradiant phase transition. We analytically solved the eigenenergies and eigenstates, and numerically simulated the system behaviors near the critical point. The critical behaviors near the superradiant phase transition are characterized by the quantum metric, defined in terms of the response of the quantum state to variation of the control parameter.  In addition, a quantum metrological protocol based on the critical behaviors of the quantum metric near the superradiant phase transition is proposed, which enables greatly  the achievable measurement precision.
	
\end{abstract}
\pacs{~}

\vskip 0.5cm

\narrowtext
\maketitle
\section{Introduction}

Superradiant phase transition (SPT) was first predicted in the Dicke model \cite{dicke1,dicke2}, which describes the collective interaction between $N$ two-level atoms (qubits) and a quantized photonic field. At the critical point where the collective coupling strength is comparable to the system frequencies, a sudden boost of the photon number occurs, as a consequence of competition between the rotating-wave and counter-rotating-wave terms in the interaction Hamiltonian. In conventional cavity quantum electrodynamics (QED) systems, the coupling-to-frequency ratio is typically much smaller than unity, so that the counter-rotating-wave terms does not produce any observable effect. Experimental realization of the SPT is challenged by this restriction, as well as by the no-go theorem--the quadratic scaling of the square of the vector potential neglected in the Dicke Hamiltonian would inhibit a sudden increase of the photon number \cite{dicke3}. These problems can be overcome by coupling the cavity mode to the transition between two electronic ground states \cite{dicke4} or two momentum states \cite{dicke5,dicke6} with the assistance of a classical field. The resulting Raman-type interaction is described by an effective Dicke Hamiltonian, where the two low-energy levels of the atoms act as the basis states of qubits.

The counter-rotating-wave terms necessary for realizing the SPT can also be effectively produced from the rotating-wave terms with a strong driving \cite{sec1}. Thus-realized effective Rabi Hamiltonian is valid in a framework rotating at the Rabi frequency of the strong driving. When viewed in the laboratory framework, the qubit undergoes a fast rotation, which makes it difficult to track the qubit state in real time, as demonstrated in a recent experiment where the SPT was realized in an on-chip effective Rabi model \cite{sec2}. In other words, the system lacks a symmetry in the laboratory framework. This difficulty can be overcome with a parametric driving applied to the photonic field, which can effectively transform the rotating-wave interaction into an asymmetric combination of both rotating-wave and counter-rotating-wave interactions \cite{sec3,sec4}. Unlike the strong-driving-based effective Rabi model, the qubits' rotation solely arises from its interaction with the quantized photonic field, and the system Hamiltonian, in the laboratory framework, still bears a ${\rm Z}_2$ symmetry, although the U(1) symmetry is broken. The SPT for such parametrically-driven system with several qubits have been theoretically explored \cite{sec5}, but in the steady state under a driving-dissipation competition, other than in the ground state of the Hamiltonian. Recently, the ground-state critical behaviors of the parametrically-driven Jaynes-Cummings model were investigated \cite{sec6}. In addition to fundamental interest, these critical phenomena can be used for realizing quantum-enhanced sensing \cite{sec7}.

We here investigate the critical behaviors of the Tavis-Cummings model \cite{thi1}, which composes of multiple qubits coupled to a quantized photonic field under a parametrical driving. Such a parametrically-driven Tavis-Cummings model possesses a ${\rm Z}_2$ symmetry in the laboratory framework, allowing the real-time tracking of the qubits’ state, as compared to the case that the drives are applied to both the qubits and the bosonic field for realizing an effective Dicke model \cite{edm}. Moreover, the infinite ratio of qubit-to-field frequency required in the Rabi model, to satisfy the thermodynamic limit so as to realize the critical phenomena, can be balanced by the number of qubits by use of the driven model. This could potentially make the practical implementation of the SPT more flexible, when the infinite frequency ratio in the parametrically-driven Jaynes-Cummings model is hard to reach. We characterize the critical phenomena around the SPT by the quantum metric \cite{thi2,thi3,thi4}, which is defined in the context of quantum information. A comparison between the nonanalytic behaviors of the quantum metric and those associated with the ground state energy is performed. We further investigate the criticality-enhanced quantum sensing based on such a model.

The paper is organized as follows. In Sec. \ref{sec2}, we describe the parameter space geometry of a given quantum system and introduce the quantum metric. In Sec. \ref{sec3}, we present the analytical derivation for the eigenstates and eigenenergies in both the normal and superradiant phases. In Sec. \ref{sec4}, we characterize the physical properties of the critical quantum system in terms of the ground state energy and the quantum metric. A dynamical quantum sensing protocol is proposed in Sec \ref{sec5} and the corresponding higher-order corrections is shown in Sec. \ref{sec6}. Sec. \ref{sec7} ends up with our conclusions. In the Appendixes, we provide some supplementary calculations for the quantum metric and the QFI.

\section{Quantum metric \label{sec2}}
In this section, we briefly review the concept of the geometry of quantum states in Hilbert space, described by the so-called quantum geometric tensor (QGT) \cite{er1,er2,er2.1}. We consider two infinitesimally separated quantum states  $\vert\phi_{k_i}\rangle$ and $\vert\phi_{k_{i+1}}\rangle$ in a parameter space $\textbf{k}=\{k_i\}$ $(i=1,2,...,M)$, where the QGT is defined as 
\begin{eqnarray}
	Q_{\mu\nu}=\langle\partial_\mu\phi\vert\partial_\nu\phi\rangle-\langle\partial_\mu\phi\vert\phi\rangle\langle\phi\vert\partial_\nu\phi\rangle=g_{\mu\nu}+iF_{\mu\nu}/2, \label{metric}
\end{eqnarray}
where $\vert\partial_j\phi\rangle$ denotes the partial derivative of $\vert\phi\rangle$ with respect to $j$ $(j=\mu, \nu)$. The real component of the QGT in (\ref{metric}) represents the quantum metric $g_{\mu\nu}={\rm Re}[Q_{\mu\nu}]$  which measures the distance between two nearby quantum states \cite{er3,er4,er6}; while the imaginary component defines the Berry curvature  $F_{\mu\nu}=2{\rm Im}[Q_{\mu\nu}]$ \cite{er77}, whose integral over a 2-dimensional closed manifold gives the first Chern number \cite{chern1,chern2,er8,er9}.

If the system is initially in a nondegenerate eigenstate $\vert\phi(k)\rangle=\vert n(k)\rangle$ whose controlled Hamiltonian $H(k)$ is slowly varied, the adiabatic theorem ensures that the system evolves remaining within the state  $\vert n(k)\rangle$. The quantum metric can be expanded as \cite{er5}
\begin{eqnarray}
	g^{(n)}_{\mu\nu}={\rm Re}\sum_{m\neq n} \frac{\langle n\vert(\partial_\mu H)\vert m\rangle\langle m\vert(\partial_\nu H)\vert n\rangle}{(E_m-E_n)^2},	 \label{gg}
\end{eqnarray}%
where $\partial_j$ is the partial derivative over $j$, and $E_r$ $(r=m, n)$ denotes the eigenvalue of the rth eigenstate. It is worth noting that the physical meaning of this quantum metric is that the larger the distance between the two states, the higher the statistical distinguishability of them. In order to investigate the SPT, we mainly focus on the ground state overlap and the correspondingly induced quantum metric. Eq. (\ref{gg}) stands out for the apparent singularities  when the energy levels occur crossings, which marks the emergence of the SPT. In order to understand the quantum metric, we hereafter treat the qubit-field coupling strength $\lambda$ and the qubit frequency $\Omega$ as the geometric parameters and fix all others, that is,  considering a family of Hamiltonians which can be smoothly controlled by the parameter $\lambda$ or $\Omega$.
\section{Quantum critical system \label{sec3}}

We here consider  multiple qubits coupled to a squeezed bosonic field \cite{sec5,san1,san3,san4}. The Hamiltonian can be described as $(\hbar=1)$
\begin{eqnarray}
	H=\omega a^{\dagger} a+\sum_{j=1}^{N}\left[\frac{\Omega_j}{2}\sigma^j_{z}+\lambda_j( a^{\dagger} \sigma^j_{-}+ a\sigma^j_{+})\right]-G( a^2+ a^{\dagger2}), \label{eq1}
\end{eqnarray}%
where $\omega$ and $\Omega_{j}$ are the frequencies of the bosonic field and of qubit $j$, respectively; $ a^{\dagger}( a)$ is the creation (annihilation) operator of the bosonic field; $\sigma^j_{x,y,z}$ are the Pauli operators and  $\sigma^j_\pm=(\sigma^j_x \pm i\sigma^j_y)/2$;  G is the squeezing coefficient and $\lambda_j$ describes the  coupling strength between qubit $j$ and the bosonic field. We then analyze the normal and superradiant phases in order to illustrate the critical behaviors of such a quantum system. 
\subsection{Normal phase}

We first discuss the critical behaviors in the normal phase. To this end, we divide $H$ into two parts,  i.e., $H=H_0+H_1$, with
\begin{eqnarray}
	H_0&=&\omega a^{\dagger} a+\sum_{j=1}^{N}\frac{\Omega_j}{2}\sigma^j_{z}-G( a^2+ a^{\dagger2}),\nonumber\\
	H_1&=&\sum_{j=1}^{N}\lambda_j( a^{\dagger} \sigma^j_{-}+ a\sigma^j_{+}).
\end{eqnarray}%

By applying a Schrieffer-Wolff transformation \cite{san2} to the Hamiltonian in Eq. (\ref{eq1}), we can remove the interaction between the qubits and the bosonic field described by  $H_1$, restricting each qubit in respective decoupled subspaces $\{\vert g\rangle_j\}$ or $\{\vert e\rangle_j\}$. The transformed Hamiltonian can then be expanded as 
\begin{eqnarray}
	H^{'}= e^{-D}He^{D}=\sum_{k=0}^{\infty}\frac{[H,D]^{(k)}}{k!},
\end{eqnarray}
where $D$ is an anti-Hermitian and block-off-diagonal operator, satisfying $[H,D]^{(k)}=[[H,D]^{(k-1)},D]$, with $[H,D]^{(0)}=H$. In order to construct $D$, we divide $H^{'}$ into the diagonal part $H_d$ and the off-diagonal part $H_o$:
{\small\begin{eqnarray}
		H_d&=&\sum_{k=0}^{\infty}\frac{1}{(2k)!}[H_0,D]^{(2k)}+\sum_{k=0}^{\infty}\frac{1}{(2k+1)!}[H_1,D]^{(2k+1)}, \nonumber\\
		H_o&=&\sum_{k=0}^{\infty}\frac{1}{(2k+1)!}[H_0,D]^{(2k+1)}+\sum_{k=0}^{\infty}\frac{1}{(2k)!}[H_1,D]^{(2k)}.
\end{eqnarray}}


Here, we require that the off-diagonal part $H_o=0$ up to the second order in $\lambda_j$, which leads to $[H_0,D]=-H_1$. With this setting, the operator $D$ can be derived as
\begin{eqnarray}
	D = \sum_{j=1}^{N}\frac{\lambda_j}{\Omega_j}(a^\dagger\sigma^j_--a\sigma^j_+)+{O}(\beta_j^{{-\frac{3}{2}}}).
\end{eqnarray} 

In the limit $\beta_j=\Omega_j/\omega\to\infty$, we can simplify the transformed Hamiltonian as  
\begin{eqnarray}
	H^{'}=\omega a^{\dagger} a+\sum_{j=1}^{N}\left(\frac{\Omega_j}{2}  \sigma^j_{z}+\frac{\lambda^2_j}{\Omega_j}a^\dagger a\sigma^j_z\right)-G( a^2+ a^{\dagger2}).
\end{eqnarray}

For simplicity but without loss of generality, we hereafter set $\lambda_j/\lambda=\Omega_j/\Omega=x_j$. With the projecting of $H^{'}$ to the ground states of the qubits, the low-energy Hamiltonian for the effective normal phase can then be written as 
\begin{eqnarray}
	H_{np}=\langle g\vert^{\otimes N} H^{'}\vert g\rangle^{\otimes N}
	=2(\alpha_{n}+G) a^{\dagger}a-G( a^2+ a^{\dagger2})-\frac{K}{2}\Omega, \label{eq7} 
\end{eqnarray}
where $\alpha_{n}=(\omega-2G)(1-g^2)/2$ with $g=\sqrt{K}\lambda/\sqrt{\Omega(\omega-2G)}$ and $K=\sum_{j=1}^{N}x_j$. To characterize the quantum metric, we here take the partial derivative of $H_{np}$ and get 
\begin{eqnarray}
	\partial_\lambda H_{np}&=&\frac{\partial H_{np}}{\partial\lambda}=\frac{-2(\omega-2G)g^2}{\lambda}a^\dagger a, \nonumber\\
	\partial_\Omega H_{np}&=&\frac{\partial H_{np}}{\partial\Omega}=\frac{(\omega-2G)g^2}{\Omega}a^\dagger a. \label{eq10}
\end{eqnarray}

At this time, the critical behaviors of the system are still disturbed by the two-photon terms. To better see the physical phenomena, we then diagonalize $H_{np}$ by introducing the field squeezed operator $S(r_{np})=e^{r_{np}(a^2-a^{\dagger2})/2}$, and choose the squeezed parameter $r_{np}=\frac{1}{4}\ln{[\alpha_{n}/(\alpha_{n}+2G)]}$ to eliminate the additional two-photon terms. Such a squeezing transformation leads to 
\begin{eqnarray}
	H^{'}_{np}=S^{\dagger}(r_{np})H_{np}S(r_{np})=\epsilon_{np}a^{\dagger}a+E_{np},
\end{eqnarray}
with the excitation energy $\epsilon_{np}=2\sqrt{\alpha_{n}(\alpha_{n}+2G)}$ and the ground-state energy $E_{np}=[\epsilon_{np}/2-(\alpha_n+G)-K\Omega/2]$. For $\omega\geq2G$, the excitation energy $\epsilon_{np}$ is real only when $g<1$, which implies that the SPT occurs at the critical point $g_c=1$. The form of phase transition parameter $g$ shows the feasibility of the SPT for finite-size few-body system, which indicates that the rigorous restraint in the coupling strength can be alleviated by increasing not only the driving strength but also the number of qubits, despite the distinct coupling strength between each qubit and the bosonic field. At this point, the low-energy eigenstates of $H_{np}$ can  be described as $\vert\phi^n_{np}\rangle=S(r_{np})\vert n\rangle\vert g\rangle^{\otimes N}$, whose corresponding eigenenergies read $E^n_{np}=n\epsilon_{np}+E_{np}$. The quantum metric for the ground state can then be calculated via Eqs. (\ref{gg}) and (\ref{eq10}) as 
\begin{eqnarray}
	g^{(0)}_{\mu\nu}={\rm Re}\sum_{n\neq 0} \frac{\langle \phi^0_{np}\vert(\partial_\mu H_{np})\vert\phi^n_{np} \rangle\langle \phi^n_{np}\vert(\partial_\nu H_{np})\vert \phi^0_{np}\rangle}{n^2\epsilon^2_{np}}, \label{eq122}
\end{eqnarray}%
where $\mu,\nu=\{\lambda,\Omega\}$. The excitation energy $\epsilon_{np}$ tends to zero when $g\to1$, meaning that the quantum metric is divergent closing to the critical point. The formula (\ref{eq7}) provides great convenience for extraction of analytical expressions of the quantum metric of the critical system. 
\subsection{Superradiant phase}
However, for $g>1$, the excitation energy $\epsilon_{np}$ becomes imaginary, indicating that the number of photons in the bosonic field becomes proportional to $\Omega/\omega$ and the field mode is occupied macroscopically so that the higher-order terms can not be neglected. By this time, the ground state has come into the superradiant phase, and the excitation energy at this point can not be described by $\epsilon_{np}$ because of the failure of the low-energy Hamiltonian. For capturing the physics of the superradiant phase accurately in this case, we displace the bosonic field mode in the original Hamiltonian $H$ of Eq. (\ref{eq1}) by using the displacement operator $D(\alpha)=e^{\alpha(a^\dagger-a)}$ $(\alpha\in R)$. Then the transformed Hamiltonian becomes
\begin{eqnarray}
	H_{\alpha}&=&D^{\dagger}(\alpha)HD(\alpha)  \nonumber\\
	&=&\omega a^\dagger a+\sum_{j=1}^{N}\left[\frac{\Omega_j}{2}\sigma^j_{z}+\alpha\lambda_j(\sigma^j_-+\sigma^j_+)\right]+\sum_{j=1}^{N}\lambda_j( a^{\dagger} \sigma^j_{-}+ a\sigma^j_{+})-G(a^2+a^{\dagger2})  \nonumber\\
	&&+\alpha(\omega-2G)(a^\dagger+a)+\alpha^2(\omega-2G).
\end{eqnarray}

To simplify the representation, we diagonalize the part of the qubits using
\begin{eqnarray}
	\vert \tilde e\rangle_j&=&\cos{\theta_j}\vert e\rangle_j+\sin{\theta_j}\vert g\rangle_j, \nonumber\\
	\vert \tilde g\rangle_j&=&-\sin{\theta_j}\vert e\rangle_j+\cos{\theta_j}\vert g\rangle_j,
\end{eqnarray}
with $\tan{2\theta_j}=2\lambda_j\alpha/\Omega_j$ and the new transition frequency $\tilde\Omega_j=\sqrt{\Omega^2_j+4\lambda^2_j\alpha^2}$. Obviously, each qubit is independent of the others and has the similar structure which enables rewriting $H_\alpha$ in the new eigenstates space as 
\begin{eqnarray}
	\tilde H_{\alpha}&=&\omega a^\dagger a+\sum_{j=1}^{N}\left[\frac{\tilde\Omega_j}{2}\tilde\sigma^j_{z}+\lambda_j\cos^2{\theta_j}( a^{\dagger} \tilde\sigma^j_{-}+ a\tilde\sigma^j_{+})\right]-\sum_{j=1}^{N}\lambda_j\sin^2{\theta_j}( a^{\dagger} \tilde\sigma^j_{+}+ a\tilde\sigma^j_{-}) \nonumber\\
	&&-G(a^2+a^{\dagger2}) +\left[\alpha(\omega-2G)+\sum_{j=1}^{N}\frac{1}{2}\lambda_j\sin{2\theta_j}\tilde\sigma^j_z\right](a+a^\dagger) +\alpha^2(\omega-2G), \label{eq11}
\end{eqnarray}
where $\tilde\sigma^j_{z,\pm}$ denote the new Pauli operators in the  eigenstates subspace $\{\vert\tilde g\rangle_j,\vert\tilde e\rangle_j\}$. To eliminate the linear terms of $a$ and $a^\dagger$ in Eq. (\ref{eq11}), we project each qubit into the new low-energy state $\vert\tilde g\rangle_j$ and choose the displacement parameter $\alpha=\pm\alpha_0=\pm\sqrt{K^2\lambda^2/4(\omega-2G)^2-\Omega^2/4\lambda^2}$. With this choice, the ratio of  the coupling strength between the rotating and counter-rotating wave terms in each qubit becomes $\gamma_j=\cos^2{\theta_j}/\sin^2{\theta_j}=(g^2+1)/(g^2-1)\to\infty$, showing that the influence of the counter-rotating wave terms can  be fairly ignored. Therefore, we simplify the Hamiltonian $\tilde H_\alpha$ to 

\begin{eqnarray}
	\tilde H_{\alpha}\simeq\omega a^\dagger a+\sum_{j=1}^{N}\left[\frac{\tilde\Omega_j}{2}\tilde\sigma^j_{z}+\tilde\lambda_j( a^{\dagger} \tilde\sigma^j_{-}+ a\tilde\sigma^j_{+})\right] -G(a^2+a^{\dagger2})+\frac{K}{4}\Omega(g^2-\frac{1}{g^2}), \label{eq12}
\end{eqnarray}
where $\tilde\Omega_j=g^2\Omega_j$ and $\tilde\lambda_j=\lambda_j(1+g^{-2})/2$ are the equivalent frequency of the corresponding qubit and the coupling strength, respectively. Note that $\tilde H_\alpha$ has the similar mathematical structure with the original Hamiltonian in Eq. (\ref{eq1}), except for the added constant term. Apparently, by applying the same method used for the normal phase, we can obtain the effective Hamiltonian for the superradiant phase as
\begin{eqnarray}
	H_{sp}=2(\alpha_{s}+G)a^\dagger a-G(a^2+a^{\dagger2})-\frac{K}{4}\Omega(g^2+\frac{1}{g^2}),   \label{eq13}
\end{eqnarray}
where $\alpha_{s}=(\omega-2G)(3g^2+1)(g^2-1)/8g^4$. Furthermore, to describe the quantum metric in the superradiant phase, we can also obtain
\begin{eqnarray}
	\partial_\lambda H_{sp}&=&\frac{\partial H_{sp}}{\partial\lambda}=\frac{(\omega-2G)(1+g^2)}{g^4\lambda}a^\dagger a, \nonumber\\
	\partial_\Omega H_{sp}&=&\frac{\partial H_{sp}}{\partial\Omega}=-\frac{(\omega-2G)(1+g^2)}{2g^4\Omega}a^\dagger a,\label{eq18}
\end{eqnarray} 
where we have discarded the terms free of $a$ and $a^\dagger$ since they do not contribute at all.
The excitation energy for the effective superradiant phase becomes $\epsilon_{sp}=2\sqrt{\alpha_{s}(\alpha_s+2G)}$, which  is real when $g>1$. Besides, the ground state energy in the superradiant phase turns to $E_{sp}=\epsilon_{sp}/2-(\alpha_s+G)-K\Omega(g^2+g^{-2})/4$. Note that the independent choices of $\alpha=\pm\alpha_0$ can only change the eigenstates but have no difference in the energy spectrum. Then the low-energy effective eigenstates in the superradiant phase can be expressed as $\vert \phi^n_{sp}\rangle=S(r_{sp})\vert n\rangle\vert \tilde g^\pm\rangle^{\otimes N}$ with a new squeezed parameter $ r_{sp}=\frac{1}{4}\ln{[\alpha_s/(\alpha_s+2G)]}$ and the reconstructed state $\vert\tilde g^\pm\rangle_j=\mp c_-\vert e\rangle_j+c_+\vert g\rangle_j$, where $c_\pm=\sqrt{(1\pm g^{-2})/2}$. The corresponding eigenenergies are $E^n_{sp}=n\epsilon_{sp}+E_{sp}$.  The quantum metric for the ground state when $g>1$ can also be obtained by substituting $\vert \phi^n_{sp}\rangle$, $\epsilon_{sp}$ and Eq. (\ref{eq18}) into Eq. (\ref{eq122}).
\section{Quantum phase transition \label{sec4}}
\begin{figure}
	\centering\includegraphics[width=10cm]{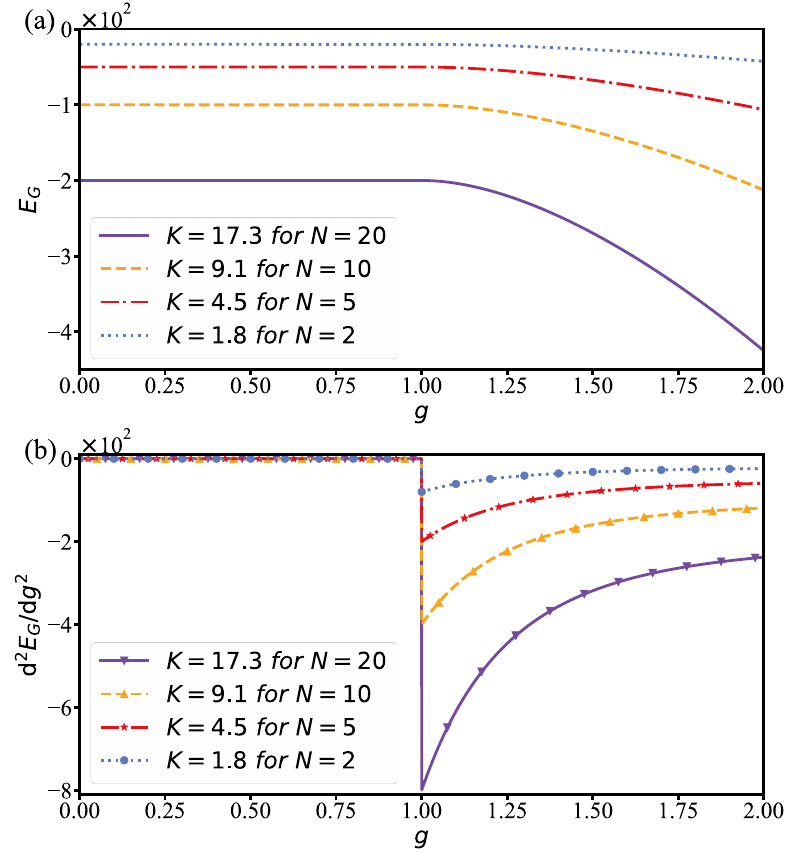}
	\caption{(a) The analytical ground state energy $E_G$ and (b) its corresponding second-order derivative ${{\rm d}^2E_G}/{{\rm d}g^2}$ as a function of the phase transition parameter $g$ with the selected number $N$ of qubits and $K$. At the critical point, the ground state energy is continuous when its second-order derivative exhibits  the apparent singularities. We here set $\Omega=20\omega$, $G=0.1\omega$.}
	\label{fig1}
\end{figure}

The analytical results shown above for the normal and superradiant phases indicate that a squeezing drive induces the finite-size few-body system to undergo a SPT at $g_c=1$. We now analyze the physical properties of the system by observing the ground state energy $E_G$ and the components of the quantum metric. We focus first on the ground state energy $E_G$ for the equivalent Hamiltonian. Note that $E_G$ is approximately equivalent to $-K\Omega/2$ for $g<1$ and becomes $-K\Omega(g^2+g^{-2})/4$ for $g>1$, meaning that the ground state energy is continuous when passing through the critical point. However, the second derivative of  $E_G$ is discontinuous at the critical point, which explains the second-order nature of the SPT, as shown in Fig. \ref{fig1}.

We now turn to the components of the quantum metric. For $g<1$, the low-energy eigenstates of the effective Hamiltonian give the expressions of the quantum metric for the ground state defined in Eq. (\ref{metric}) with 
\begin{eqnarray}
	g^{(0)}_{\lambda\lambda}&=&\frac{(\omega-2G)G^2g^2K}{8\Omega\alpha^2_{n}(\alpha_n+2G)^2},\nonumber\\
	g^{(0)}_{\Omega\Omega}&=&\frac{(\omega-2G)^2G^2g^4}{32\Omega^2\alpha^2_n(\alpha_n+2G)^2},\nonumber\\
	g^{(0)}_{\lambda\Omega}&=&g^{(0)}_{\Omega\lambda}=-\frac{(\omega-2G)^{\frac{3}{2}}G^2g^3K^{\frac{1}{2}}}{16\Omega^{\frac{3}{2}}\alpha^{2}_n(\alpha_n+2G)^2}. \label{eq15}
\end{eqnarray}

\begin{figure}
	\centering\includegraphics[width=10cm]{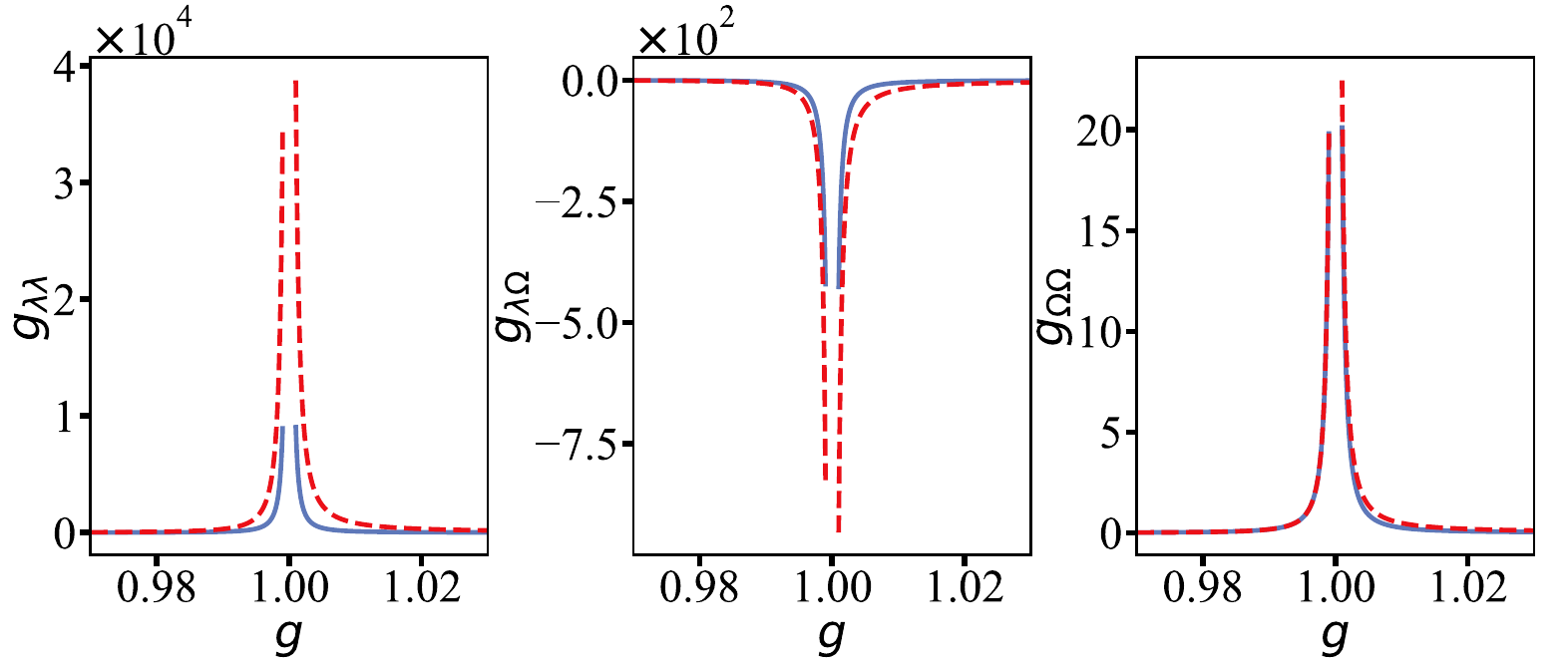}
	\caption{The solution of the quantum metric components under the effective Hamiltonian $H_{np}$ and $H_{sp}$ when $K=4.5$ for $N=5$ (blue solid line) and $K=17.3$ for $N=20$ (red dashed line). Each component of the quantum metric exhibits a divergent behavior when $g$ approaches to the critical point $g_c$. We here set $\Omega=20\omega$, $G=0.1\omega$. }
	\label{fig2}
\end{figure}

The divergent behavior emerging in $g^{(0)}_{\mu\nu}$ $(\mu, \nu = \lambda, \Omega)$ of (\ref{eq15}) also demonstrates the emergence of the SPT as the system is close to the critical point $g_c$ (see Appendix \ref{A} for the detailed derivation), as well depicted in Fig. \ref{fig2}.  The situation is also true for $g>1$, as these components for the ground state are similarly
divergent according to
\begin{eqnarray}
	g^{(0)}_{\lambda\lambda}&=&\frac{(\omega-2G)G^2(1+g^2)^2K}{32g^{10}\Omega\alpha^2_s(\alpha_s+2G)^2}+\frac{(1+3g^2)NK}{8g^6\Omega(1+g^2)\alpha_{s}},\nonumber\\
	g^{(0)}_{\Omega\Omega}&=&\frac{(\omega-2G)^2G^2(1+g^2)^2}{128g^8\Omega^2\alpha^2_s(\alpha_s+2G)^2}+\frac{(\omega-2G)(1+3g^2)N}{32g^4\Omega^2(1+g^2)\alpha_s}, \nonumber\\
	g^{(0)}_{\lambda\Omega}&=&g^{(0)}_{\Omega\lambda}=-\frac{(\omega-2G)^{\frac{3}{2}}G^2(1+g^2)^2K^{\frac{1}{2}}}{64g^9\Omega^{\frac{3}{2}}\alpha^2_s(\alpha_s+2G)^2}    -\frac{(\omega-2G)^{\frac{1}{2}}(1+3g^2)NK^{\frac{1}{2}}}{16g^5\Omega^{\frac{3}{2}}(1+g^2)\alpha_s}. \label{eq16}
\end{eqnarray}
Overall, the critical behavior characterized and exhibited by the ground state energy as well as by the quantum metric when $g$ approaches $g_c$, demonstrates the occurrence of the SPT.

\section{Quantum sensing and encoding scheme \label{sec5}}
With this quantum critical system, the quantum Fisher information (QFI) can be obtained and is expected to exhibit  a divergent behavior when the system is approaching the critical point, indicating that the criticality-enhanced quantum sensing \cite{liu1.1,liu1.2,liu1.3,liu1.4,liu1.5,liu1.6} may potentially be allowed. The QFI, defined by the quantum Cram\'{e}r-Rao bound \cite{clml}, shows the absolute lower bound of the measurement pricision for an input state and is independent of the measurement methods \cite{liu1}. Such a lower bound indicates that the QFI as large as possible is required so as to get the higher measurement precision when the relevant parameter is estimated. Recently, a dynamic framework was proposed \cite{liu2} to realize the quantum sensing without the particular requirement for state preparation or adiabatic evolution. As was illustrated in Ref. \cite{liu2}, the Hamiltonian of Eqs. (\ref{eq7}) and (\ref{eq13}) both satisfy the relation \replaced{$[H_{\alpha'},\Lambda]=\sqrt{\Delta}\Lambda$}{$[H_\alpha,\Lambda]=\sqrt{\Delta}\Lambda$} with \replaced{$\Delta=16\alpha'(\alpha'+2G)$}{$\Delta=16\alpha(\alpha+2G)$}, where \replaced{$\alpha'$}{$\alpha$} is equal to   $\alpha_{n}$ for $g<1$ and becomes $\alpha_{s}$ for $g>1$, enabling prominent improvement in the achievable measurement precision (see Appendix \ref{B}).

We here show the ultimate precision about the parameter \replaced{$\alpha'$}{$\alpha$}, expressed by the QFI as

\begin{eqnarray}
	\mathcal{F}_{\alpha'}(t)\simeq1024G^2(\alpha'+2G)^2 \times\frac{[\sin(\sqrt{\Delta}t)-\sqrt{\Delta}t]^2}{\Delta^3}{\rm Var}[\hat {P}^2]_{\left\vert \varphi\right\rangle}, \label{111}
\end{eqnarray}%
where $P=i(a^\dagger-a)/\sqrt{2}$ is the quadrature of the bosonic field with the initial state $\vert\varphi\rangle$. The nonanalytic behaviors at $\Delta\to0$ enable enhancing greatly the accuracy of the relevant parameter in both the normal and superradiant phases, allowing the quantum sensing by encoding the physical quantity related to \replaced{$\alpha'$}{$\alpha$}. Note that the QFI is expected to be equivalent to the inverted variance of the measurement, meaning the potentially good performance of the sensing in virtue of the specific encoding scheme. 
\begin{figure}
	\centering\includegraphics[width=10cm]{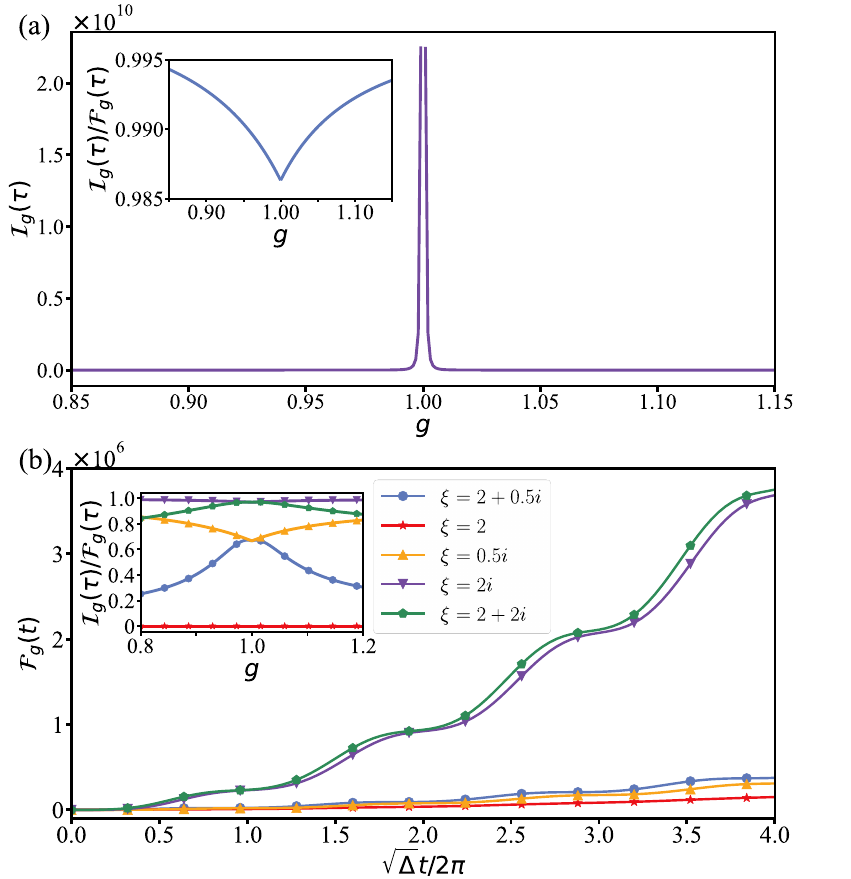}
	\caption{(a) Quantum sensing by homodyne detection of the bosonic field. The inverted variance $\mathcal{I}_g(\tau)$ as a function of the phase transition parameter $g$. Inset: The ratio $\mathcal{I}_g(\tau)/\mathcal{F}_g(\tau)$ reaches its local maximum with the evolution time $\tau=2\pi/\sqrt{\Delta}$. We here set $\xi=3i$. (b) QFI $\mathcal{F}_g(t)$ as a function of the evolution time $t$ for the different coherent parameter $\xi$ with $g=0.96$. Inset: The different ratio $\mathcal{I}_g(\tau)/\mathcal{F}_g(\tau)$ for the different coherent state. }
	\label{fig4}                                                                                                  
\end{figure}

We initialize the bosonic field in the coherent state $\vert\varphi\rangle=\vert\xi\rangle$ and all qubits in their ground states. To quantify the precision of the parameter estimation, we define the inverted variance \replaced{$\mathcal{I}_{\alpha'}(t)=\chi ^2_{\alpha'}(t)/(\Delta{X})^2$}{$\mathcal{I}_\alpha(t)=\chi ^2_\alpha(t)/(\Delta{X})^2$}, where \replaced{$\chi_{\alpha'}(t)=\partial_{\alpha'}\langle X\rangle_t $}{$\chi_\alpha(t)=\partial_\alpha\langle X\rangle_t $} is the susceptibility of the observable $\langle X\rangle_t$ and $(\Delta X)^2$  is the variance of the other  quadrature $X=(a+a^\dagger)/\sqrt{2}$ of the bosonic field. The mean value and variance of $X$ can easily be obtained as  
\begin{eqnarray}
	\langle X\rangle_t&=&\sqrt{2}\xi_r\cos{(\sqrt{\Delta}t/2)} -4\sqrt{2}\xi_i(\alpha'+2G)\Delta^{-\frac{1}{2}}\sin{(\sqrt{\Delta}t/2)}, \nonumber\\
	(\Delta X)^2&=&\frac{1}{2}\cos^2{(\sqrt{\Delta}t/2)}+8(\alpha'+2G)^2\Delta^{-1}\sin^2{(\sqrt{\Delta}t/2)},
\end{eqnarray}%
where $\xi_r$ and $\xi_i$ represent the real and imaginary component of $\xi$, respectively.  The high-precision quantum metrology requires the  inverted variance to have the same order when compared to the QFI. We can then obtain the local maximum of the inverted variance after an evolution time  $\tau_n=2 n\pi/\sqrt{\Delta}$ $(n\in{N}^+)$:
\begin{eqnarray}
	\mathcal{I}_{\alpha'}(\tau_n)&=& 4096\xi^2_i(\alpha'+G)^2(\alpha'+2G)^2\Delta^{-2}\tau^2_n.
\end{eqnarray}%

Clearly, the inverted variance is independent of the real component of $\xi$ and the number of qubits. We further get the inverted variance of the relative parameter $A$ as \replaced{$\mathcal{I}_A(\tau_n)=(\partial\alpha'/\partial A)^2\mathcal{I}_{\alpha'}(\tau_n)$}{$\mathcal{I}_A(\tau_n)=(\partial\alpha/\partial A)^2\mathcal{I}_\alpha(\tau_n)$}, which exhibits the same critical behaviors as $\Delta\to0$.  As shown in Fig. \ref{fig4}(a), the inverted variance $\mathcal{I}_g(t)$ is of the same order to the QFI in both the normal and  superradiant phases, indicating the potential application of such a system for quantum-enhanced sensing. Notice that, the arbitrary dynamics evolution accounting for the divergent behavior
for both $\mathcal{I}_g$ and $\mathcal{F}_g$ at specific time points, is applicable to arbitrary pure or mixed state form with coherent states as the basis states.

Moreover, we go one step further to show  the comparison with different coherent coefficient $\xi$. As shown in Fig. \ref{fig4}(b), the imaginary component of $\xi$ plays a  crucial role in  improvement of  the relevant parameter precision; however, the real component of $\xi$ has a negative influence on such a measurement scheme. The cause lies in the fact that $\xi_i$ enhances the QFI but has no effect on the inverted variance. Additionally, as will be readily seen that as long as $\xi$ is chosen appropriately, the higher precision will be obtained when compared to the encoding scheme proposed in Ref. \cite{sec7}, where the bosonic field is initially in one of the Fock states.

Compared to the adiabatic sensing protocol proposed with quantum Rabi model \cite{liu1.3}, our calculations prove that such a dynamical protocol can achieve the Heisenberg limit, not resorting to the long operation time to satisfy the adiabatic condition. This thus relaxes the requirement when referring to the practical implementation of the protocol. Moreover, for the estimation of the bosonic field frequency, the conventional Ramsey interferometric protocol may also achieve the optimal Heisenberg-scaling precision; however, the preparation of a specific initial state would be a challenge.

\begin{figure}
	\centering\includegraphics[width=10cm]{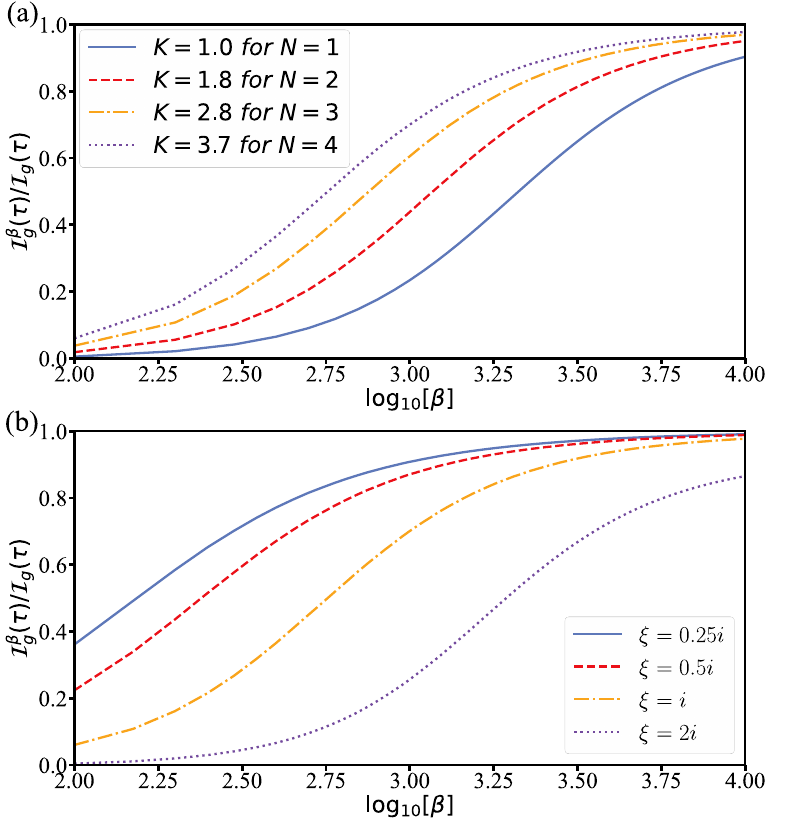}
	\caption{Ratio $\mathcal{I}^{\beta}_g(\tau)/\mathcal{I}_g(\tau)$ versus $\beta$ (in ${\rm log}_{10}[\cdot]$) for (a) the selected number $N$ of qubits and $K$ and (b) the different coherent parameter $\xi$.  The requirement of the infinite frequency ratio $\beta$ can be relaxed provided that, the number $N$ of the qubits increases, or, the coherent parameter $\xi$ decreases. We here set $G=0.1\omega$ and $g=0.96$.  }
	\label{fig5}                                                                                                  
\end{figure}

\section{Higher-order corrections \label{sec6}}

The analysis shown above is in the limit $\beta=\Omega/\omega\to\infty$, which is necessary for the occurrence of the SPT. We here show a higher-order  correction by considering the imperfect frequency ratio of the qubits to  the bosonic field. For this purpose, we expand the transformed operator $D$ as
\begin{eqnarray}
	D^r&=&D+\sum_{j=1}^{N}\left[\frac{\lambda_j\omega}{\Omega^2_j}(a^\dagger\sigma^j_--a\sigma^j_+)+\frac{2\lambda_jG}{\Omega^2_j}(a^\dagger\sigma^j_+-a\sigma^j_-)\right] \nonumber\\
	&&+\sum_{j=1}^{N}\frac{4\lambda^3_j}{3\Omega^3_j}(aa^\dagger a\sigma^j_+-a^\dagger aa^\dagger \sigma^j_-)+{O}(\beta^{{-\frac{5}{2}}}),
\end{eqnarray}
and decouple the Hamiltonian in Eq. (\ref{eq1}), keep $H'$ up to the fourth order term in $\lambda_j/\Omega_j$, project it to $\mathcal{H}_\downarrow$, and obtain
{\small \begin{eqnarray}
		H^r_{np}=H_{np}-\frac{\omega(\omega-2G)g^2N}{K\Omega}a^\dagger a+\frac{G(\omega-2G)g^2N}{K\Omega}(a^2+a^{\dagger2}) +\frac{(\omega-2G)^2g^4}{K\Omega}a^\dagger aa^\dagger a.   \label{eq14}
\end{eqnarray}}

It shows that the leading-order correction to $H_{np}$ is equivalent to a quartic potential for the bosonic field, with the coefficient inversely proportional to the number of qubits, indicating that the lack of the infinite frequency ratio in some extent can be compensated by the increase of the number of qubits. This result is verified  by the numerical outcomes shown in Fig. \ref{fig5}(a). In addition,  the leading-order correction can also be influenced by the alteration of the coherent parameter $\xi$. As Fig. \ref{fig5}(b) shows, in a way,  the imperfect frequency ratio can be relaxed with the choice of a smaller pure imaginary $\xi$. In other words, there exists a trade-off as  improvement in the inverted variance by increasing $\xi$ means strengthening the requirement for the frequency ratio.

The physical model could be implemented in kinds of spin-boson systems. For a circuit quantum electrodynamics architecture, the interaction between multiple superconducting qubits and a resonator is naturally described by the Tavis-Cummings model, while the squeezing driving can be realized by a Josephson-junction-coupled nonlinear process \cite{apply1,apply2}. In an ion trap, such an interaction model can be achieved by the coupling between the internal and external degrees of freedom of multiple trapped ions, mediated by a laser tuned to the first red sideband, and the squeezing driving can be realized with a Raman-type driving \cite{apply3}.

\section{Conclusion \label{sec7}}
In conclusion, we have presented the critical behaviors of multiple qubits coupled to a bosonic field with a squeezed drive. The introduction of a parametric drive induces an effective Dicke superradiance transition, without resorting to the strong qubit-field coupling, providing a new method to engineer the exotic transition in quantum systems. We have analytically solved the eigenenergies and eigenstates, and characterized the second-order SPT by using the ground state energy and the quantum metric. We have further proposed a criticality-enhanced quantum sensing scheme by making use of the system's dynamics, where the obtained inverted variance is on the same order when compared to the QFI, exhibits a divergent scaling approaching the critical point. With the developed homodyne detection method used commonly, for instance, in circuit QED \cite{measurement}, the inverted variance, introduced to quantify the precision of the parameter estimation, can be reached. We found that the requirement for the infinite ratio of qubit-to-field frequency to realize the quantum sensing protocol can be relaxed by increasing the number of qubits, enhancing the feasibility of the explored case when referring to the practical implementation in experiment.

\appendix\section{The analytical derivation of the quantum metric} \label{A}
We here show the detailed derivation of the quantum metric in the normal and superradiant phases. For $g<1$, the low-energy eigenstate of the effective Hamiltonian $H_{np}$ can be expressed as $\vert\phi^0_{np}\rangle=S(r_{np})\vert 0\rangle\vert g\rangle^{\otimes N}$. To calculate the quantum metric defined in Eq. (\ref{metric}), we get 
\begin{eqnarray}
	\frac{\partial\vert\phi^0_{np}\rangle}{\partial\lambda}&=&-\frac{G(\omega-2G)^{\frac{1}{2}}gK^{\frac{1}{2}}}{4\Omega^{\frac{1}{2}}\alpha_n(\alpha_n+2G)}(a^2-a^{\dagger 2})S(r_{np})\vert 0\rangle\vert g\rangle^{\otimes N}, \nonumber\\
	\frac{\partial\vert\phi^0_{np}\rangle}{\partial\Omega}&=&\frac{G(\omega-2G)g^2}{8\Omega\alpha_n(\alpha_n+2G)}(a^2-a^{\dagger 2})S(r_{np})\vert 0\rangle\vert g\rangle^{\otimes N}.
\end{eqnarray}
The quantum metric can then be expressed as $	g_{\mu\nu}={\rm Re}[Q_{\mu\nu}]$, where $\mu,\nu=\{\lambda,\Omega\}$ (see Eqs. (\ref{metric}) and (\ref{eq15}) in the main text).

For $g>1$, the low-energy eigenstate is $\vert\phi^0_{sp}\rangle=S(r_{sp})\vert 0\rangle\vert \tilde g^\pm\rangle^{\otimes N}$, where $\vert \tilde g^\pm\rangle_j=\frac{1}{\sqrt{2}}(\mp\sqrt{1- g^{-2}}\vert e\rangle_j+\sqrt{1+ g^{-2}}\vert g\rangle_j)$. Then  we have
\begin{eqnarray}
	\frac{\partial\vert\phi^0_{sp}\rangle}{\partial\lambda}&=& \frac{G(\omega-2G)^{\frac{1}{2}}(g^2+1)K^{\frac{1}{2}}}{8\Omega^{\frac{1}{2}}g^5\alpha_s(\alpha_s+2G)}(a^2-a^{\dagger2})\vert\phi^0_{sp}\rangle - \sum_{j}\frac{\sqrt{2}K^{\frac{1}{2}}}{2(\omega-2G)^{\frac{1}{2}}\Omega^{\frac{1}{2}}g^3}S(r_{sp})\vert0\rangle \nonumber\\
	&&\otimes [\pm\frac{1}{\sqrt{1- g^{-2}}}\vert e\rangle_j+\frac{1}{\sqrt{1+ g^{-2}}}\vert g\rangle_j]\otimes  \vert \tilde g^\pm\rangle^ {\otimes N-1}       ,     \nonumber\\
	\frac{\partial\vert\phi^0_{sp}\rangle}{\partial\Omega}&=&-\frac{G(\omega-2G)(g^2+1)}{16\Omega g^4\alpha_s(\alpha_s+2G)}(a^2-a^{\dagger2})\vert\phi^0_{sp}\rangle+ \sum_{j}\frac{\sqrt{2}}{4\Omega g^2}S(r_{sp})\vert0\rangle\otimes \vert \tilde g^\pm\rangle^ {\otimes N-1} \nonumber\\
	&&\otimes [\pm\frac{1}{\sqrt{1- g^{-2}}}\vert e\rangle_j+\frac{1}{\sqrt{1+ g^{-2}}}\vert g\rangle_j]   .
\end{eqnarray}
The quantum metric in the superradiant phase can also be described as $	g_{\mu\nu}={\rm Re}[Q_{\mu\nu}]$ (see Eqs. (\ref{metric}) and (\ref{eq16}) in the main text).
\section{The detailed derivation of the QFI} \label{B}

The effective Hamiltonian in the normal phase can be written as 
\begin{eqnarray}
	H_{np}&=&2(\alpha_{n}+G) a^{\dagger}a-G( a^2+ a^{\dagger2})-\frac{K}{2}\Omega 
	= 2GP^2+\alpha_n(X^2+P^2)+Const, \label{B1}
\end{eqnarray}
where $\alpha_{n}=(\omega-2G)(1-g^2)/2$. By choosing $H^{'}_0=2GP^2$ and  $H^{'}_1=X^2+P^2$, we get 
\begin{eqnarray}
	\hat A &=& -i[H^{'}_0,H^{'}_1]=-4G(XP+PX),\nonumber\\
	\hat B &=& -[H_{np},[H^{'}_0,H^{'}_1]]
	=-16G\alpha_nX^2+16G(\alpha_n+2G)P^2.
\end{eqnarray}
Then, we simplify the expression \replaced{$[H_{\alpha'},\hat\Lambda]=\sqrt{\Delta}\hat\Lambda$}{$[H_\alpha,\hat\Lambda]=\sqrt{\Delta}\hat\Lambda$} as
\begin{eqnarray}
	[H_{np},[H_{np},[H^{'}_0,H^{'}_1]]]=\Delta[H^{'}_0,H^{'}_1],
\end{eqnarray}
where $\hat \Lambda=i\sqrt{\Delta}\hat A-\hat B$, and find that Eq. (\ref{B1}) satisfies the equation with $\Delta=16\alpha_{n}(\alpha_{n}+2G)$. Next, we express the transformed local generator  \replaced{$h_{\alpha'}=-i(\partial_{\alpha'} U^\dagger_{\alpha'})U_{\alpha'}=i U^\dagger_{\alpha'}(\partial_{\alpha'} U_{\alpha'})$}{$h_\alpha=-i(\partial_\alpha U^\dagger_\alpha)U_\alpha=i U^\dagger_\alpha(\partial_\alpha U_\alpha)$} using $\hat A$ and $\hat B$ as
\begin{eqnarray}
	h_{\alpha'}&=&-i\sum_{k=0}^{\infty}\frac{(it)^{k+1}}{(k+1)!}[\hat H_{\alpha'},\hat H^{'}_1]_k
	=\hat H^{'}_1t+\frac{\cos{(\sqrt\Delta t)}-1}{\Delta}\hat A-\frac{\sin{(\sqrt\Delta t)}-\sqrt\Delta t}{\Delta^{\frac{3}{2}}}\hat B. \label{B5}
\end{eqnarray}
Apparently, as $g\to g_c$, the third term in Eq. (\ref{B5}) makes the dominant role, and  the QFI can then be given as
\begin{eqnarray}
	\mathcal{F}_{\alpha'}&=&4{\rm Var}[h_{\alpha'}]_{\vert\Phi\rangle}
	\simeq1024G^2(\alpha_{n}+2G)^2 \frac{[\sin{(\sqrt\Delta t)}-\sqrt\Delta t]^2}{\Delta^3}{\rm Var}[\hat P^2]_{\vert\Phi\rangle}.
\end{eqnarray}
The situation is similar for $g>1$.

\section{The finite-size scaling of the measurement precision enhancement} \label{C}
We here describe a finite-size scaling for the improvement of the measurement precision. We simplify the Hamiltonian in Eq. (\ref{eq1}) as 
\begin{eqnarray}
	H = \omega a^\dagger a+ \Omega J_z +\frac{\lambda}{\sqrt{N}}(a^\dagger J_-+aJ_+)-G(a^2+a^{\dagger2}), \label{eq34}
\end{eqnarray}
where $J_{m}=\sum_{j=1}^{N}{\frac{1}{2}\sigma_{m}}$ $ (m=x,y,z)$ are the collective qubit operators and $J_{\pm}=J_x\pm iJ_y$ denote the collective raising and lowering operators with pseudospin $j=N/2$.
\begin{figure}
	\centering\includegraphics[width=10cm]{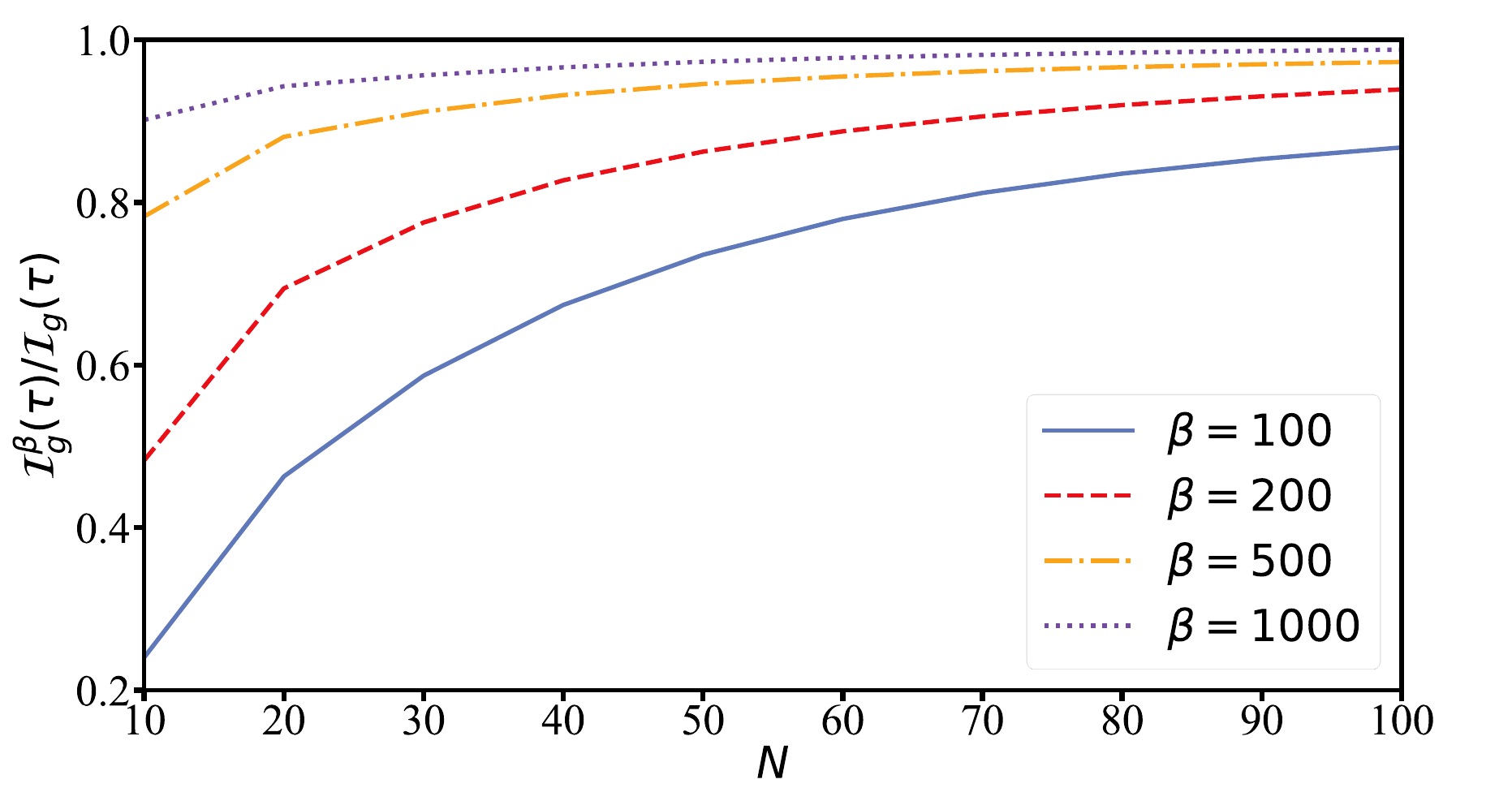}
	\caption{Ratio $\mathcal{I}^{\beta}_g(\tau)/\mathcal{I}_g(\tau)$ versus the number $N$ of qubits for the different frequency ratio $\beta$. We here set $\xi=i$ and $g=0.96$.  }
	\label{fig6}                                                                                                
\end{figure}

We first simplify the whole Hamiltonian by use of the Holstein-Primakoff transformation, with the defined collective operators that are expressed with the newly introduced bosonic operators $b$ and $b^\dagger$:  
\begin{eqnarray}
	J_z &=& b^\dagger b-j, \nonumber\\
	J_+ &=& b^\dagger\sqrt{2j-b^\dagger b},\nonumber\\
	J_- &=& \sqrt{2j-b^\dagger b}b.   \label{eq35}
\end{eqnarray}
We then substitute  Eq. (\ref{eq35}) into Eq. (\ref{eq34}) and get 
\begin{eqnarray}
	H = \omega a^\dagger a+ \Omega(b^\dagger b-j)+\frac{\lambda}{\sqrt{N}}(a^\dagger\sqrt{2j-b^\dagger b}b+a b^\dagger\sqrt{2j-b^\dagger b})-G(a^2+a^{\dagger2}),\label{eq36} 
\end{eqnarray}
based on which the numerical simulation  is performed. In Fig. \ref{fig6}, we show the ratio $\mathcal{I}^{\beta}_g(\tau)/\mathcal{I}_g(\tau)$ as a function of the number $N$ of qubits for several frequency ratio $\beta$. It clearly shows that, $\mathcal{I}^{\beta}_g(\tau)/\mathcal{I}_g(\tau)$ becomes larger with the increase of $N$, and tends to unity when $N$ approaches infinity, i.e., the thermodynamic limit,  for all the cases with different $\beta$. In practice, maybe it's uneasy to get a larger $\beta$. In such a case, a large $N$ can be a compensation. In other words, the optimal ratio Ratio $\mathcal{I}^{\beta}_g(\tau)/\mathcal{I}_g(\tau)$ can be obtained through selecting a trade-off between $\beta$ and $N$. 
\section{The acquirement of the inverted variance} \label{D}
To realize the enhanced parameter sensitivity in practical implementation, we need the inverted variance, which is defined as $\mathcal{I}_g=\partial^2_g\langle X\rangle_t/(\Delta{X})^2$, to approach the QFI $\mathcal{F}_g$ as close as possible. To quantize this, we require to do the measurement of the average field quadrature $\langle X\rangle_t$ at a specific time point for the dynamical evolution, by presetting a system parameter. This can be achieved by following the conventional homodyne detection method used commonly, for instance, in circuit QED (please see \cite{measurement} as a review). Along with this, the derivation of  $\langle X\rangle_t$ with respect to the modulated parameters $g$, i.e., $\partial_g\langle X\rangle_t$ can be obtained. To get the variance $\Delta{X}$, we can repeat multiple $(j=1,2,…,N)$ measurements for such a parameter $g$, get the mean value $\overline{\langle X\rangle}$, and gain $\Delta X=\sqrt{\sum_{j=1}^{N}(\langle X\rangle_j-\overline{\langle X\rangle})/N}$.

\section*{Funding}
National Natural Science Foundation of China ( 12274080, 11874114, 11875108); National Youth Science Foundation of China (12204105); Educational Research Project for Young and Middle-aged Teachers of Fujian Province (JAT210041); Natural Science Foundation of Fujian Province (2021J01574, 2022J05116).
\section*{Disclosures}
The authors declare that there are no conflicts of interest related to this article.

\section*{Data availability}
Data underlying the results presented in this paper are not publicly available at this time but may be obtained from the authors upon reasonable request.
\bibliography{reference}

\end{document}